# MULTI-DIRECTIONAL FLOW AS TOUCH-STONE TO ASSESS MODELS OF PEDESTRIAN DYNAMICS

**Author:** Tobias Kretz
**Affiliation:** PTV Group
**Address:** Haid-und-Neu-Straße 15, D-76131 Karlsruhe, Germany
**Email:** Tobias.Kretz@ptvgroup.com
**Phone:** +49 721 96 51-7280
**Fax:** +49 721 96 51-693

## ABSTRACT

For simulation models of pedestrian dynamics there are always the issues of calibration and validation. These are usually done by comparing measured properties of the dynamics found in observation, experiments and simulation in certain scenarios. For this the scenarios first need to be sensitive to parameter changes of a particular model or – if models are compared – differences between models. Second it is helpful if the exhibited differences can be expressed in quantities which are as simple as possible ideally a single number. Such a scenario is proposed in this contribution together with evaluation measures. In an example evaluation of a particular model it is shown that the proposed evaluation measures are very sensitive to parameter changes and therefore summarize differences effects of parameter changes and differences between models efficiently, sometimes in a single number. It is shown how the symmetry which exists in the achiral geometry of the proposed example scenario is broken in particular simulation runs exhibiting chiral dynamics, while in the statistics of 1,000 simulation runs there is a symmetry between left- and right-chiral dynamics. In the course of the symmetry breaking differences between models and parameter settings are amplified which is the origin of the high sensitivity of the scenario against parameter changes.

## INTRODUCTION

Simulations for engineering purposes need verification, validation, and calibration. For the simulation of pedestrians basic verification tests are for example maintaining a given speed or not walking through walls Compilations of such tests are for example (Brunner, et al. 2009) and (International Maritime Organization 2007).

Concerning calibration it is usually assumed that it is most important to reproduce the fundamental diagram of pedestrian dynamics. Different empirical studies have resulted in very different pedestrian fundamental diagrams (Schadschneider, et al. 2009) such that recent studies were restricted to one-dimensional movement (Seyfried, et al. 2005), which discovered cultural differences (Chattaraj, Seyfried und Chakroborty 2009) and resolved the dynamics in high detail (Portz und Seyfried 2011). Other investigated properties include the fundamental diagram in a corridor in combination with the flow through bottlenecks (Liddle, Seyfried und Steffen 2011) and the merging flow in T junctions (Zhang, Klingsch, et al. 2011). Such scenarios are necessary in a calibration process such that a simulation tool is enabled to faithfully assess if the available capacity meets the demand and to accurately measure travel and delay times. However, these scenarios have a strong emphasize on how pedestrians adjust speeds on density when everyone is walking in about the same direction. Directional choices on the contrary are of minor relevance in these settings or to be more precise: these settings bare only minor potential to resolve differences between models or parameter settings regarding walking directions. In the aforementioned one-dimensional (single-file) movement experiments direction choices even have been deliberately excluded entirely. However, maybe with the exception of emergency egress simulations in the field of fire safety engineering, such situations are not the only settings that occur in reality. If one limits oneself to these in the calibration process, one also limits a model in its range of applications. Examples where multi-directional flows are predominant would be station halls or city squares.

An obvious step beyond situations where all pedestrians essentially go into the same direction – and where therefore direction choices are more relevant – was done early when bi-directional flow situations were investigated with simulations (Blue und Adler 1999). The famous phenomenon of lane-formation (Hoogendoorn und Daamen 2005) (Kretz, Grünebohm, et al. 2006) (Zhang, Klingsch, et al. 2012) was observed (Helbing und Molnar 1997) in the original formulation of the Social Force Model (Helbing und Molnar 1995) and also in a Cellular Automaton model (Burstedde, et al. 2001). Model extensions have been proposed to improve the pedestrian agents' performance in counterflow by a "commoving dynamic potential" (Kretz, Kaufman und Schreckenberg 2008) or an "anticipation floor field" (Suma, Yanagisawa und Nishinari 2012) (Nowak und Schadschneider 2012).

Compared to situations where all pedestrians walk into (approximately) the same direction in counterflow situations the number of lanes per unit width, band index (Yamori 1998) and the stability of lanes, merging or split frequencies are all additional evaluation measures which summarize the state of the system to a few numbers and which can be used to compare empirical data with simulation results.

However, counterflow situations still have just one primary axis of movement. Pedestrian movement can also be multi-directional. In multi-directional movement walking is characterized at least as much by the choice of a walking direction as by the choice of a walking speed in response to the locations and velocities of the surrounding pedestrians.

It is therefore desirable to have at least one multi-directional scenario included in the calibration process. Typically situations which hold multi-directional flows in reality are rather complex and arbitrary in the sense that there is no prototypical variant from which all others can be derived – compare the aforementioned stations and city squares. However, a scenario that serves for validation and calibration should have a low complexity such that it can be modeled quickly and reproduced reliably by different modelers and with different tools. At the same time it should be possible to define simple measures with which the system dynamics (the walking behavior) can be quantified – comparable to the band index in counterflow situations. Finally it would be welcome, if the scenario amplifies differences resulting from differences of the computation of the walking direction such that the differences can be seen with the naked eye in an animation of the simulation.

Already calling for low complexity suggests having a highly symmetric geometry, but there is also a second reason for this: Some models are formulated with a small degree of spatial symmetry. Cellular Automata models with a square grid have two symmetry axes (Kretz und Schreckenberg 2007) and with a triangular grid (hexagonal cells) they have six (Gipps und Marksjö 1985). Also data structures with a similarly low degree of symmetry are often used when models are implemented as software which in principle have a higher degree of symmetry. In both cases it is desirable to know the effects of the low(er) degree of symmetry (i.e. the distinct axes of discretization). This can best be seen in geometries which have a high degree of symmetry and settings where many different walking directions occur. One would expect to get the high degree of symmetry replicated in the simulation results, at least in the statistics of many repeated runs (with different random numbers). Again settings with just one or two predominantly occurring walking directions cannot resolve such (non-) isotropy issues.

Finally it should be possible – at least in principle – to actually implement the setting of the calibration example as experiment. A condition like "straight corridor with periodic boundaries" excludes the possibility for experiment and observation.

## SCENARIO DESCRIPTION

72 pedestrians are set at the same time (normally simulation start t=0.0 s) exactly or approximately equally distributed on a circle line (i.e. one pedestrian each 5 degree), see FIGURE 1. Their desired (or preferred or maximum) walking speeds are either fixed to 5.4 km/h (1.5 m/s), chosen to be in a small range as for example between 5.39 and 5.41 km/h or – if a model requires only allows one of a discrete set of desired speeds in which 1.5 m/s is not included – to a value or range which is close to it. Then each pedestrian has to walk to the exactly opposite area (opposite to the starting), see FIGURE 2.

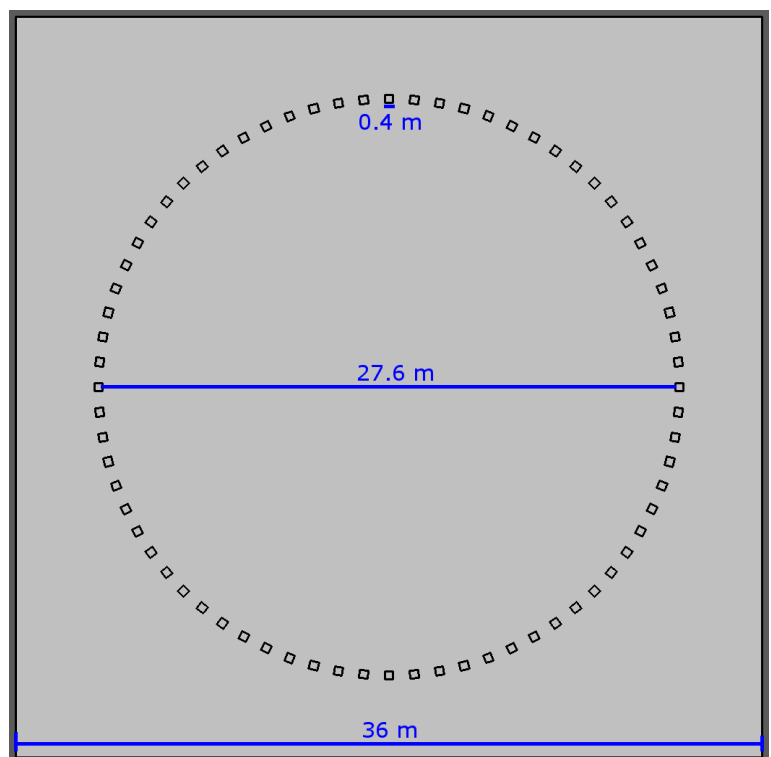

**FIGURE 1: Walking area (36 x 36 m²) and starting locations, here starting areas with an extent of 0.4 x 0.4 m² are used. At the beginning one pedestrian per starting area is set into the simulation. As starting position on the starting area either the center is chosen or the position on the area is chosen randomly.**

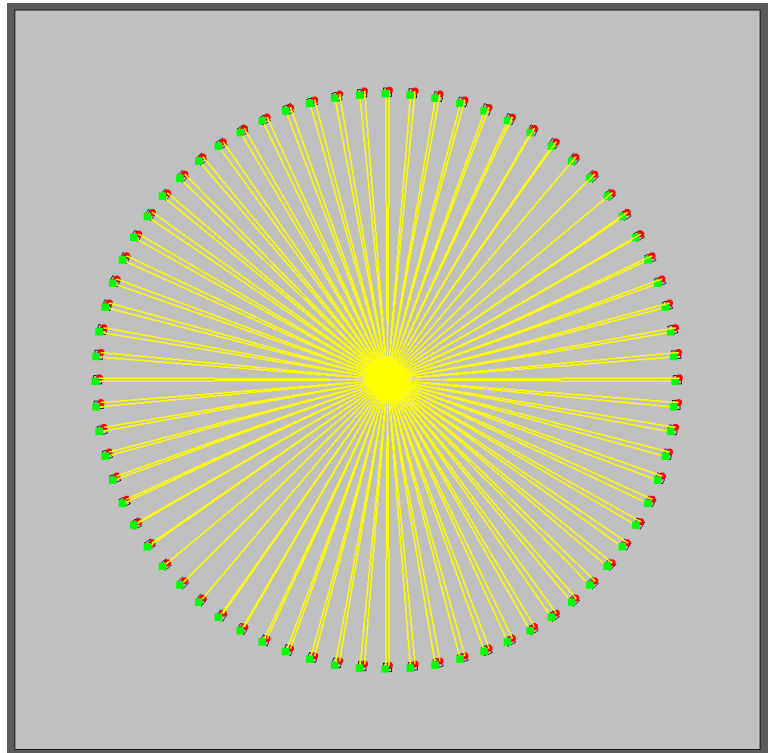

**FIGURE 2: Pedestrians' routes (in the sense of OD relations); a red dot marks a starting area. It is linked with a yellow line with a green dot which marks the corresponding destination area, which is another pedestrian's starting area.**

## EXISTING WORK

This is not the first time that such a scenario is investigated. Preceding works include (van den Berg, Lin und Manocha 2008)[1], (Guy, et al. 2010)[2], and (Ondrej, et al. 2010)[3]. These works do not give numerical statistic results, but keep to the visual impression of the animation of the simulation of this scenario. It is indeed remarkable, how strongly – compared to one- or bi-directional flow in a corridor – this scenario distinguishes different models already for the naked eye.

## EVALUATION MEASURES

To define meaningful evaluation measures one has to consider all possible ways the simulation could evolve in any model which produces even just remotely realistic results of pedestrian dynamics.

Hypothetically the OD movement as indicated in FIGURE 2 could be accomplished if every pedestrian stayed at the same distance from the center and all 72 pedestrians did an angular clockwise or counter-clockwise movement until everyone reaches the oppositely located destination. However, this would not be very realistic as one can assume that reducing the radius at least a little bit would allow finishing the process earlier. This hypothetical resolution is therefore considered to be not relevant in the following.

---

[1] See also: http://youtu.be/1Fn3Mz6f5xA
[2] See also http://youtu.be/hpYdjHzHTkY
[3] See also: http://youtu.be/586qhaDwr24

It is instead assumed that at first the simulated pedestrians will walk (approximately) toward the center point (inbound movement phase). At some time they will cover a minimum area extent – the convex hull encompassing all pedestrians will have smallest area extent in the course of the simulation. The time when this happens should approximately coincide with the time when the largest distance between the center point and a pedestrian becomes minimal and with the point in time when the average walking speed becomes minimal. Following this state a simulation model than will either keep the system in this state forever, i.e. not resolve the situation, or it will resolve the situation in one way or another.

If the situation is not resolved at all the simulated pedestrians will not arrive at their destination, i.e. the simulation time becomes infinite. If the situation is resolved the average time of arrival as well as the time of arrival of the last pedestrian is a first and simple measure to distinguish different simulation models or different parameter sets of one and the same model.

Of particular interest is, however, how exactly the situation is resolved, which means to have a close look at what happens at the aforementioned time of highest density or smallest walking speed. Exclusively looking at the state of the system at one particular simulation time step might veil relevant information in noise. It is therefore beneficial to average all evaluation measures over a time span which includes a few simulation time steps. A time of 3 seconds is a value where an average can be built even for the models with rather large simulation time steps (the author is aware of no model with a time step larger than one second) and one can expect that within a time span of 3 seconds the situation does not change so much that one would actually be faced with two different system states.

The following three evaluation measures are proposed to extract and condense relevant information of a simulation run of the scenario described above. The first two should be given with their average over a time interval which usually includes the time of highest density (and/or lowest average walking speed). In repeated simulations this point in time will vary, nevertheless the same time interval should be used for all runs. As a consequence the interval might only "usually" include that point in time.

First: Level of Service (LOS). This is an established evaluation measure. Applying the LOS concept implies that one has to define a set of level breakpoints which are to be used. In the example evaluation below the scheme by Weidmann works well to provide insight. In other result data sets other schemes might be more helpful. Obviously when two LOS maps are compared the same scheme needs to be applied.

Second, a measure $A$ is defined specifically to evaluate this particular scenario. For measure $A$ the whole walking area is divided into – or covered with – four areas which all cover exactly half of the area: „north", „east", „south", and „west", where „north" covers the area of all positive y coordinates and so on. The idea of this division is to aggregate measurements on these areas, therefore they are called "measurement areas" furthermore. Each of these measurement areas overlaps with each of its two halves with another measurement area. On each measurement area the orientation vectors $\boldsymbol{o}$ of all pedestrians are documented for all time steps in the above described 3 seconds interval. The orientation vector $\boldsymbol{o}$ is the unitless velocity vector normalized to a length of 1. For the four measurement areas over said time interval the average orientation vectors $\boldsymbol{o}^{north}$, $\boldsymbol{o}^{east}$, $\boldsymbol{o}^{south}$, and $\boldsymbol{o}^{west}$ are measured. With the $x$ and $y$ components $o_x$ and $o_y$ of these vectors measure $A$ is defined as

$A = o_x^{south} - o_x^{north} + o_y^{east} - o_y^{west}.$

Expanded to a formulation with velocity components this writes as

$$A = \frac{1}{N_t} \sum_{t=t_0}^{t_1} \left( \frac{1}{P_s} \sum_p^{P_s} \frac{1}{\sqrt{v_{px}^2 + v_{py}^2}} v_{px} - \frac{1}{P_n} \sum_p^{P_n} \frac{1}{\sqrt{v_{px}^2 + v_{py}^2}} v_{px} + \frac{1}{P_e} \sum_p^{P_e} \frac{1}{\sqrt{v_{px}^2 + v_{py}^2}} v_{py} - \frac{1}{P_w} \sum_p^{P_w} \frac{1}{\sqrt{v_{px}^2 + v_{py}^2}} v_{py} \right)$$

$N_t$ is the number of simulation time steps within the relevant 3 seconds interval, $t_0$ and $t_1$ are the beginning and the end of the time interval. $P_s$ is the number of pedestrians on the southern measurement area, $P_n$, $P_e$, and $P_w$ accordingly the numbers on the other measurement areas. $v_{px}$ and $v_{py}$ are the velocity components of pedestrian $p$. The inner sums all run over those pedestrians who are located on the corresponding measurement areas: if a sum runs from 1 to $P_s$ this means that it runs over all pedestrians located on the southern measurement area.

On the north and south (east and west) measurement area the dominant movement is in positive or negative $y$ ($x$) direction. So $A$ is built from the non-dominant movement directions, which would average – in the average over all pedestrians – to zero for bee-line movement. If $A$ is clearly positive (negative), one is faced with a left-turning (right-turning) movement. As all four components at maximum can have an absolute value of 1, $A$ can take values between -4 and 4.

Third, measure $P$ is defined specifically for the evaluation of this particular scenario: compute for every pedestrian $i$ the average orientation vector $o^i$ and the average position vector $p^i$ throughout his existence in the simulation, i.e. from the start until he reaches his destination and from these the $z$ coordinate of the vector product: $z = p^i_x\, o^i_y - p^i_y\, o^i_x$. If this value is positive then – if the pedestrian did not do some zigzagging movement – the pedestrian has had the center of the walking area on his left hand side i.e. he did a left turning movement. In this way – with the sign of $z$ – each pedestrian in a simulation run can be assigned to a rotation direction and – with the absolute value $|z|$ - there is also a measure how pronounced the arc was which the pedestrian has walked. Measure $P$ then simply is the sum of the $z$ values of all 72 pedestrians.

Expanded to a formulation with velocity components this writes as:

$$P = \sum_{p=1}^{72} \left( \left( \frac{1}{T_p} \sum^{T_p} x_p \right) \left( \frac{1}{T_p} \sum^{T_p} \frac{1}{\sqrt{v_{px}^2 + v_{py}^2}} v_{py} \right) - \left( \frac{1}{T_p} \sum^{T_p} y_p \right) \left( \frac{1}{T_p} \sum^{T_p} \frac{1}{\sqrt{v_{px}^2 + v_{py}^2}} v_{px} \right) \right)$$

where $T_p$ is the number of time steps pedestrian $p$ has been in the simulation before arriving at the destination, $x_p$ and $y_p$ are the location coordinates of pedestrian $p$ and $v_{px}$ and $v_{py}$ the velocity components of pedestrian $p$. The latter four variables are to be understood at a particular simulation time step.

The difference between the two measures $A$ and $P$ is that measure $A$ is area-oriented and measure $P$ is pedestrian-oriented. The advantage of measure $P$ over measure $A$ is that it does not rely on the choice of a measurement time interval as for all pedestrians each position they have had at a time during the simulation enters the calculation of measure $P$. Second, measure $P$ does not rely on having to split up the area to separate measurement areas. The advantage of $P$ can also be interpreted as its disadvantage: it does not resolve what happens around the moment of highest density as measure $A$ does. In total one can expect that $P$ and $A$ are correlated. Measure $A$ has no physical dimension while measure $P$ has the dimension of a distance.

## EXAMPLE EVALUATION

The usefulness of the above proposed and defined evaluation measures can best be demonstrated by demonstrating the whole process with a concrete example. For this the pedestrian simulation software PTV Viswalk (Kretz, Hengst und Vortisch 2008)in version 5.40 has been chosen.

To demonstrate the ability of the evaluation measures to distinguish between different models (or parameter settings it has been chosen to vary the two parameters $g$ and $h$ of the *dynamic potential* (or *map of estimated remaining travel time*) method. The method and with it the two parameters by design have a strong impact on the walking direction choices of the simulated pedestrians. Aside parameters $g$ and $h$ and the desired speeds – which were equally distributed between 5.39 and 5.41 km/h – the value of no other parameter has been varied and the defaults have been used in all simulations. The following few paragraphs give a short summary of the dynamic model utilized in PTV Viswalk and the dynamic potential extension.

The pedestrian model applied in Viswalk is an elaborate version of the *Social Force Model* (Helbing und Molnar 1995), very similar to the elliptical specification II as described in (Johansson, Helbing und Shukla 2007). In the Social Force Model and all its variants for each pedestrian in each time step an acceleration vector is computed. This acceleration results as a sum from forces "pulling" the pedestrian towards his destination, (repulsive) forces from static obstacles, and (usually repulsive) forces from other pedestrians. While in the original formulation of the Social Force Model only the velocity of the acting pedestrian (the source of the force) was considered and in another variant the inter-pedestrian force even depended only on the distance (Helbing, Farkas und Vicsek 2000), elliptical specification II considers in the repulsive inter-pedestrian forces the relative velocity between the acting and sensing pedestrian and the relative velocity related to the translation vector between both pedestrians. This combination can be interpreted to implicitly model anticipation in the form of linear extrapolation of the other's movement.

In addition to the refinement of the formulation of the inter-pedestrian forces, the Social Force Model for Viswalk was extended further with a dynamic potential approach to compute the direction of the desired velocities of the pedestrians. The dynamic potential is a map of estimated remaining travel time to destination for each point of a grid from which the destination can be reached. The map is recalculated each time step, which is 0.1 s.

The computation process can be summarized as follows: each time step an estimated walking time to pass small (usually 20 x 20 cm²) grid cells covering as a map all of the walking area is assigned to these grid cells. The estimated walking times for these small pieces (the cells) are summed up starting at the destination to result in a field of estimated remaining travel time from each grid cell to the destination. The summation process is accomplished by solving the Eikonal Equation numerically. (Kimmel und Sethian 1998). The negative gradients in the field of estimated remaining travel times are used as direction of the desired velocity of a pedestrian located at the corresponding spot. Details of this method have been published previously (Kretz, Große, et al. 2011) (Kretz, Hengst und Perez Arias, et al. 2012) (Kretz 2012) , repeating them here is beyond the scope of this contribution. Relevant to know is the role of the two parameters *g* and *h* which control the numerical details:

$$\frac{1}{f(x,y)} = 1 + \max\left(0, g\left(1 + h\frac{\vec{v}(x,y)\cdot\nabla S(x,y)}{\langle v^0\rangle|\nabla S(x,y)|}\right)\right)$$

This equation gives what is used as right hand side of the Eikonal Equation for locations which are occupied by a pedestrian. *S* is the map of distances to destination, $<v^0>$ is the average of desired speeds of pedestrians who are heading for that destination, and *v* is the velocity of the location occupying pedestrian. If a location is unoccupied then *f=1*.

Parameter *g* sets the basic impact the presence of a pedestrian has on the estimated delay time it takes to move from one coordinate to a coordinate which is opposite of this particular pedestrian. If parameter *g* has a value *g=0.0* then the map of estimated remaining travel time to destination essentially is a map of distance to destination (*f=1* everywhere), which obviously implies that the map does not have to be recalculated each time step. With *g=0.0* the method is no longer an extension of the Social Force Model and its elliptical specification II, but makes pedestrians navigate as originally proposed. Typical values for parameter *g* range up to *g=3.0*. For a discussion of parameter ranges see (Kretz, Große, et al. 2011).

The value of parameter *h* determines how much the walking direction of a pedestrian is considered, when the estimated walking time delay he imposes is calculated. With a value *h=1.0* the value doubles compared to base strength if the pedestrian walks with free speed exactly away from the destination and it vanishes when he walks toward it. If on the contrary *h=0.0* then the walking direction and speed have no effect at all. If parameter *g=0.0* then the value of parameter *h* has no effect.

The major gain from integrating a dynamic potential approach into a model of pedestrian dynamics is that it smoothly interpolates between collision avoidance on short distances and – which is the prime motivation for usage – elaborate steering around more remote high density regions, i.e. avoiding already situations where collision avoidance or conflict resolution become relevant. This is not possible in models which have a hard cut-off distance for influences. The desire to avoid high density regions is automatically balanced with the extra path length a pedestrian has to do on a potential detour. The cost of the method lies in the increased computation time.

The geometry of the proposed evaluation scenario was created using the "rotate network" and "read additionally" functionality of Viswalk. In this way the symmetric creation of the geometry could be easily accomplished. Four measurement areas were created which each cover one of the above described hemispheres; a fifth included all of the walking area. The measured properties included number of pedestrians, world coordinates, velocity components, and speed.

As a first step the time until the last pedestrian has reached his destination is computed as an average of 1,000 simulation runs for different value combinations of parameters *g* and *h,* see TABLE 1.

|   |   | Value of parameter *h* | | | | |
|---|---|---|---|---|---|---|
|   |   | **0.0** | **0.5** | **1.0** | **1.5** | **2.0** |
|   | **0.0** | 42.52 | 42.52 | 42.52 | 42.52 | 42.52 |
|   | **0.5** | 30.70 | 29.99 | 29.40 | 29.10 | 29.09 |
| ***g*** | **1.0** | 28.72 | 28.52 | 28.41 | 28.54 | 28.68 |
|   | **1.5** | 29.11 | 28.97 | 28.76 | 28.84 | 28.89 |
|   | **2.0** | 29.53 | 29.47 | 29.26 | 29.18 | 29.07 |
|   | **3.0** | 30.20 | 30.10 | 29.88 | 29.71 | 29.53 |

**TABLE 1: Time [s] until the last pedestrian has reached his destination. Averages of 1,000 simulation runs each. Within these 1,000 runs the smallest measured value was typically 2 s smaller, the largest was 3 s larger. The standard deviations are at about 0.7 s. All values are slightly larger if *g=0.0*.**

The results in TABLE 1show that the pedestrians manage to conclude the process clearly quicker when *g>0.0* compared to *g=0.0*. Compared to the difference between *g=0.0* and *g=0.5* further changes at the value of g have only minor effect. However, it can be seen that the time does not decrease with increasing values of g, but that there is a minimum at *g=1.0* (and *h=1.0*).

As – compared to the static case with *g=0.0* – the parameter combination *g=1.0* and *h=1.0* appears to pronounce the effect of the method of the dynamic potential most, the following detailed evaluations are done as a comparison between the cases *g=0.0* and *g=1.0 / h=1.0*.

As the time to finish the process drops when the method of the dynamic potential is activated by giving parameter *g* a positive value, the method achieves what is suggested when the dynamic potential is interpreted as "map of estimated remaining travel time" and pedestrians are said to desire to walk into the direction of estimated earliest arrival. However, how exactly is it achieved that the process is finished earlier? How do the pedestrians walk? In which way(s) do the simulations with a positive value for parameter *g* differ from the simulations with *g=0.0*?

The first answer is that the pedestrians do not close up so much in the central region. The Level of Service (LOS) plots as shown in FIGURE 3 demonstrate this clearly.

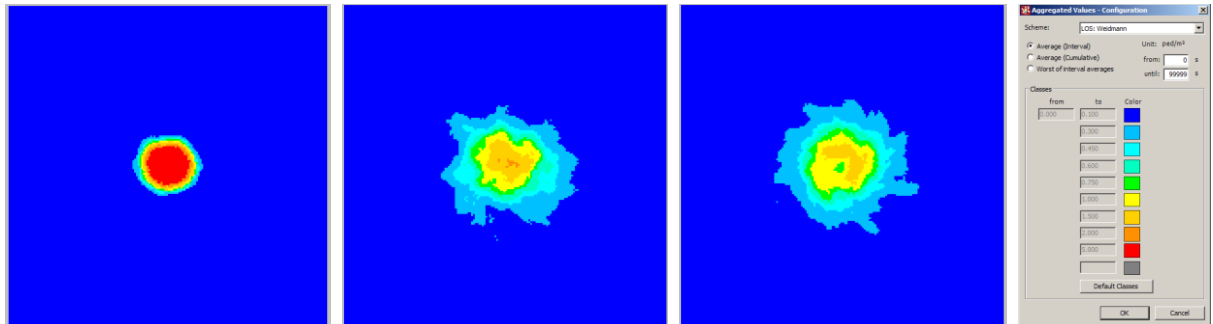

**FIGURE 3: LOS plots and color coding. All plots show the averages of seconds 10.0 to 20.0 of the corresponding simulations. The LOS value was computed according to the scheme of (Weidmann 1993) which is displayed in the rightmost figure (d). The figure to the left (a) resulted from a *g=0.0* simulation. The two figures in the middle (b) and (c) resulted from two simulation runs with *g=1.0* and *h=1.0* whose settings only differed from each other in the seed value for the random number generator. Note that the difference between both is that in figure (b) the highest density is shown in the center, while in (c) it is in a ring around the center.**

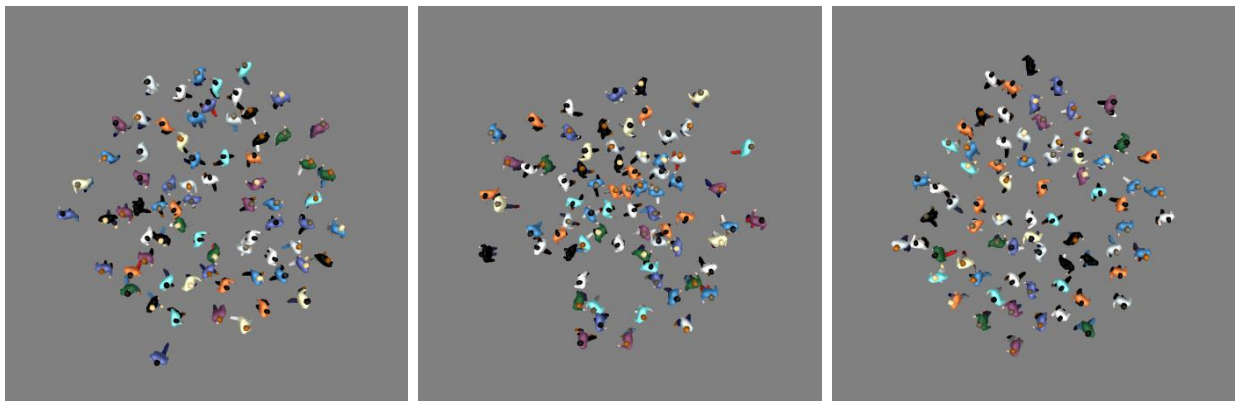

**FIGURE 4: a) left turning, b) symmetric, c) right turning. The screenshots all were made at simulation time 14 s. Looking closely the movement directions can be seen in the still image, nevertheless movies give a much better impression, see part I of http://youtu.be/Ivbstw8FIuo. FIGURE 3c) belongs to the same simulation run as a) and FIGURE 3b) to the one of b). It appears therefore that with pronounced rotational movement the highest density does not occur in the very center, but in a ring surrounding it.**

Looking at the visualizations of various runs with *g=1.0* and *h=1.0* and varying the random number generator's seed value gives more insight: in most simulation runs the pedestrians self-organize with a clearly visible rotational movement. In some cases a left turn, in some cases a right turn. There are, however, also runs in which no clear turning direction can be seen. All three cases are depicted in FIGURE 4. If *g=0.0* the rotational movement never occurs.

Following this insight the degree of the rotational movement can be quantified with measures *A* and *P*.

As relevant time interval for measure *A* (if g=1.0 and h=1.0) the interval [12.0..14.9] has been identified. In this time interval the average value of the velocity component $v_x$ ($v_y$) changes its sign on the eastern and western (northern and southern) measurement areas which means that the movement changes from inbound to outbound.

FIGURE 5 shows the histogram for measure *A*. It shows two peaks for relatively large absolute values of *A*. One could argue that there is a third and smaller peak at moderate negative values, but if at all it is not very pronounced. Assuming that there is no third peak at or near zero it can be concluded that the case of small absolute values of *A* is not a state in itself, but in such simulation runs the rotation movement is just weakly pronounced.

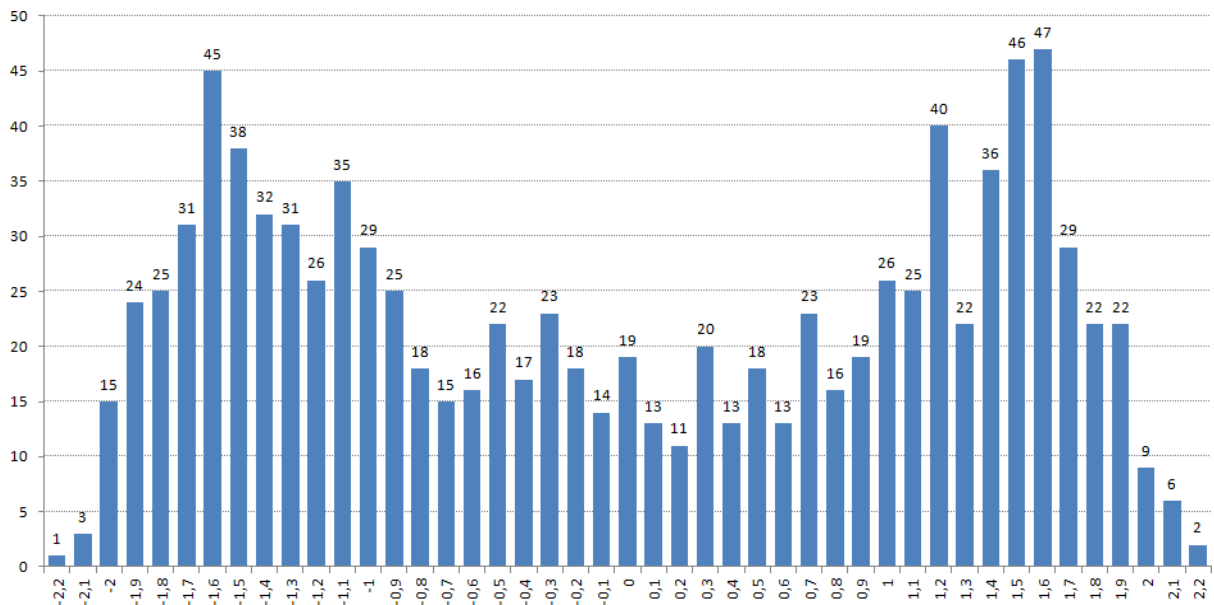

**FIGURE 5: Frequency distribution of measure *A* for 1,000 simulation runs with *g=1.0* and *h=1.0*.**

In 1,000 simulation runs (*g=1.0* and *h=1.0*) the minimum and maximum values found for measure *A* were -2.1813 and 2.1844 and the average was 0.136 with a standard deviation of 1.2746. In 514 simulation runs there was *A<0*. This implies that the null hypothesis that there is no side preference or some other bias in the simulation cannot be rejected for any required significance level of 16.4% or smaller. To reject the null hypothesis on a significance level of 1% a ratio of at least 537:463 was needed.

In 1,000 simulation runs the minimum for *P* was found to be -74.45 and the maximum 75.45 with an average of -0.14 at a standard deviation of 40.03. *P<0.0* was found in 506 simulation runs. This is even closer to the expectation value of 499.5 than the according result for measure *A*.

In 485 simulation runs the majority of pedestrians had *z>0.0* and in 508 runs the majority had z<0.0 which leaves 7 cases where equally many had positive and negative *z* values.

Measures *A* and *P* are closely correlated as FIGURE 6 shows. There are, however, cases in which measures *A* and *P* have a different sign, which means that the two measures disagree on the observed rotation direction. Looking at such simulation runs often reveals different rotation directions side-by-side, i.e. multiple lanes with opposite walking (and rotating) directions. In these and other cases with small absolute values of *A* and *P* the simulation runs look anything but similar to runs with parameter *g=0.0* where the absolute values for *A* and *P* are also small, see FIGURE 8.

There is an – albeit weak – anti-correlation of about -0.21 of the absolute values of *A* and *P* with the average walking times and the walking times of the last pedestrian to arrive (see FIGURE 7). This means that while even with small absolute values of *A* and *P* the pedestrians walk more efficient than with simulations with parameter *g=0.0* it is – on average – more efficient to walk with a pronounced rotational movement.

In a histogram of measure *P* there cannot be seen a peak at moderate negative values. So it is concluded that the third peak which could be recognized in the histogram of measure A (see FIGURE 5) is a statistical fluctuation within the presented 1,000 simulation runs and that it does not point to the existence of some effect.

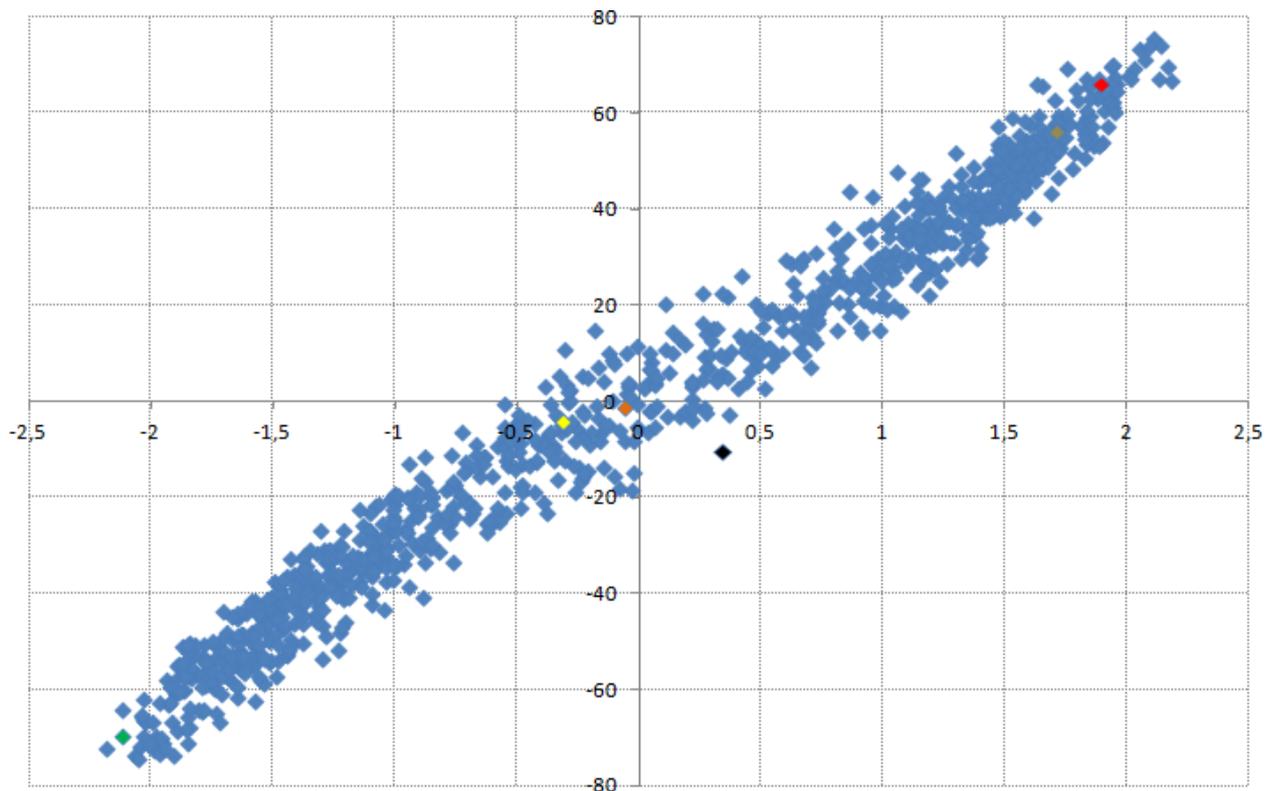

**FIGURE 6: Measure *P* (y axis, [m]) plotted vs. measure *A* (x axis). Some values are highlighted with a special color: red, orange, and green correspond to the runs of figures 4a), b), and c) respectively. Brown marks a run with comparatively pronounced rotation, but rather large travel times, while yellow marks a run with opposite properties: weakly pronounced rotation, but small travel times. Black is the run in which the two measures *A* and *P* gave the most diverging estimation on the rotation direction. The brown, yellow, and black marked runs can be seen in the second part of http://youtu.be/Ivbstw8FIuo.**

It is possible in Viswalk to explicitly activate a side preference. 1,000 simulation runs were done with a preference to evade and pass on the right side. A preference for evading and passing to the right side should result in a predominant left-turning movement. And this is what actually could be observed as measure *A* in 568 out of 1,000 runs was positive, indicating a dominant left turning movement, compare FIGURE 8. The ratio 568:432 rejects the null hypothesis that there is no direction bias on a significance level of $10^{-5}$. The average value of *A* shifted to 0.214, while the minimum and maximum values observed with [-2.133..2.138] remained very similar to the ones of the base case discussed above.

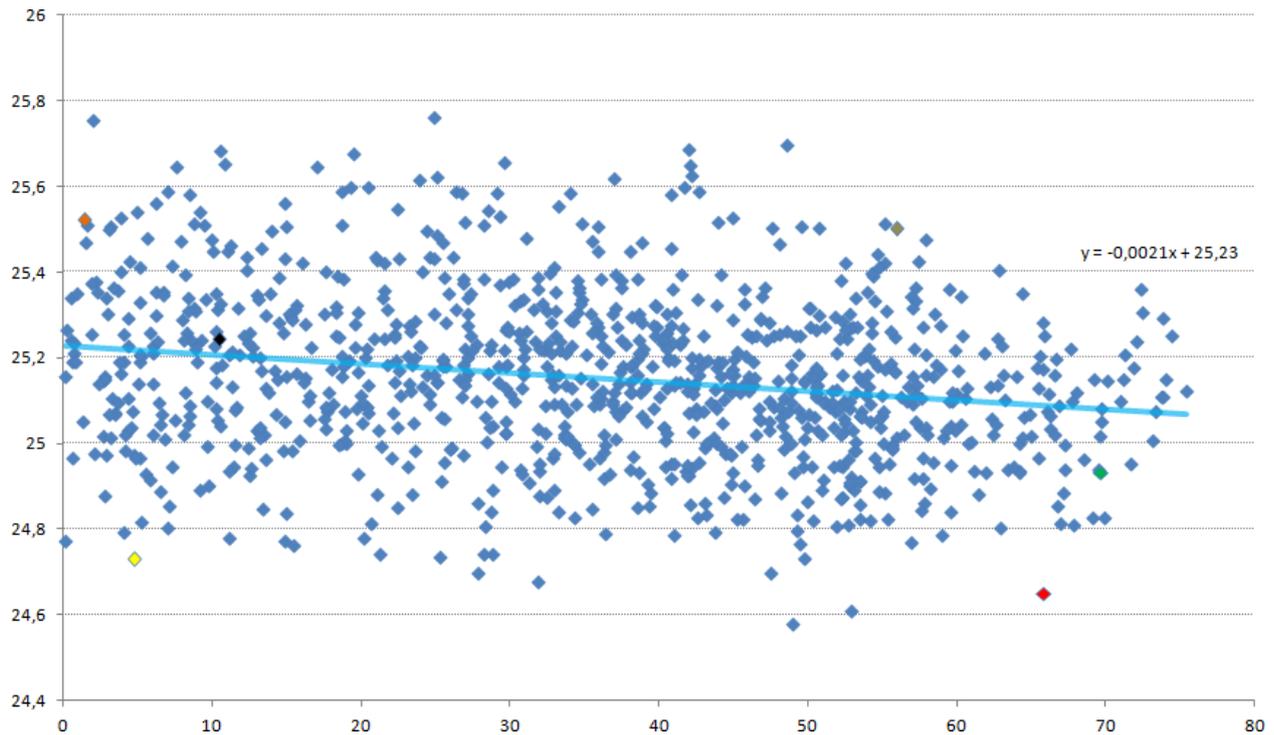

**FIGURE 7: Average of travel time [s] in each run vs. measure _P_. Colors of selected values as described in caption of FIGURE 6.**

It is interesting to relate the results of the simulations with _g=1.0_ and _h=1.0_ to those of simulations with _g=1.0_ but _h=0.0_, i.e. when there is no walking direction dependence in the calculation of estimated delays stemming from a particular person, and with the case _g=0.0_, i.e. pedestrians desiring to walk into the direction of the shortest path. FIGURE 8 shows the histograms for measure _A_ of these two and the other two previously discussed variants (_g=1.0_ and _h=1.0_ without and with side preference). It can be seen that both, with _g=1.0 / h=0.0_ and with _g=0.0_ there are not two peaks to the left and the right, but that there is only one peak which is much more pronounced for _g=0.0_. From this it follows that it is mainly the model element controlled with parameter _h_ which is responsible for the rotational movement. However, as TABLE 1shows, most of the reduction of travel times can be achieved already with _g>0.5_ and _h=0.0._ No or only small rotational movement is necessary for this. The reason can already be seen in the LOS plots of FIGURE 3: with _g>0.0_ the pedestrians never form a crowd as compact as with _g=0.0_, no matter what the value of parameter _h_ is. As animation this can be seen in the third part of http://youtu.be/Ivbstw8FIuo. In a sense therefore parameter _g_ controls the radial movement, while parameter _h_ controls the angular movement. At least in this scenario the effect of the parameters appears to be mostly decoupled which is helpful for calibrating their values from empirical data.

## CONCLUSIONS AND OUTLOOK

It could be shown that the proposed scenario is capable to produce various movement patterns already by stochastic fluctuations, i.e. when all model parameters are kept and only different random numbers are applied in the simulation. The movement patterns can easily be analyzed and can be presented and compared in compact form.

As parameter changes have a clear effect on the proposed evaluation measures it can be assumed that there is an at least comparable impact on the evaluation measures when the proposed scenario is simulated with different models of pedestrian dynamics. The scenario is therefore capable to distinguish various models as well as different parameter settings of these models from each other. In case that one day there is empirical data on this scenario available, it should be well suited to assess the quality of models and parameter settings precisely.

The scenario appears to be a good test case to unveil spatial or directional biases in a simulation tool as well as a low degree of symmetry in the model formulation or the implementation of the model to a computer program. Applying any simulation tool to the scenario therefore is a chance for the model to fail in these aspects – falsification instead of verification. To the extent of the investigations presented here, Viswalk did not "take the chance". All results were compatible with the assumption that there is no direction bias, except for the case when the bias was introduced intentionally. In this case the bias could be clearly identified. As a consequence of Viswalk passing all tests the potential of the proposed scenario to unveil low degrees of symmetry in a model or its implementation could not be demonstrated explicitly in this contribution.

Concerning the rotational movement it would be interesting to investigate further, if the rotation direction results from a spontaneous symmetry breaking (Brading und Castellani 2008) (Tanedo 2011), which means that the symmetry is broken from exactly symmetric initial conditions. It could after all be the case that the rotation direction is pre-determined in each run by the slight deviations from perfect symmetry which already exist in the simulation setup (the initial conditions) like the exact position on the input area, small variations in desired speed. As a third possibility the symmetry breaking could be a consequence of the limited precision of computers as such, i.e. that a simulation with initial conditions which are perfectly symmetric to the computing system's precision still would exhibit symmetry breaking, but an analytic solution of the dynamic equations would not.

In an experiment walking at least the distribution of free walking speeds can be expected to be wider than assumed in this contribution. So contrary to an investigation of the symmetry breaking properties where it is interesting to have every pedestrian at the same desired speed for comparison with empirical data it is interesting to re-run the scenario with wider sped distributions and quantify their impact.

The results given in this work are just a fraction of what could be measured and used for comparison. One could for example measure the time evolution of density, particularly the peak value, in a small disc-shaped area around the center or the time evolution of the average speed.

Also measure $P$ has not been exploited fully. As it is the sum of the $z$ values the information on the distribution of these values is lost. While two simulation runs can have identical $P$ value, they might differ in their statistics of $z$ values. A variant for $P$ would be to build the $z$ values not as product of orientation and position, but velocity and position. Then $P$ would be a measure for the average angular momentum of a pedestrian. Finally it is also possible to investigate a time-dependent $P(t)$, respectively $z(t)$ instead of computing these as averages of the whole simulation run.

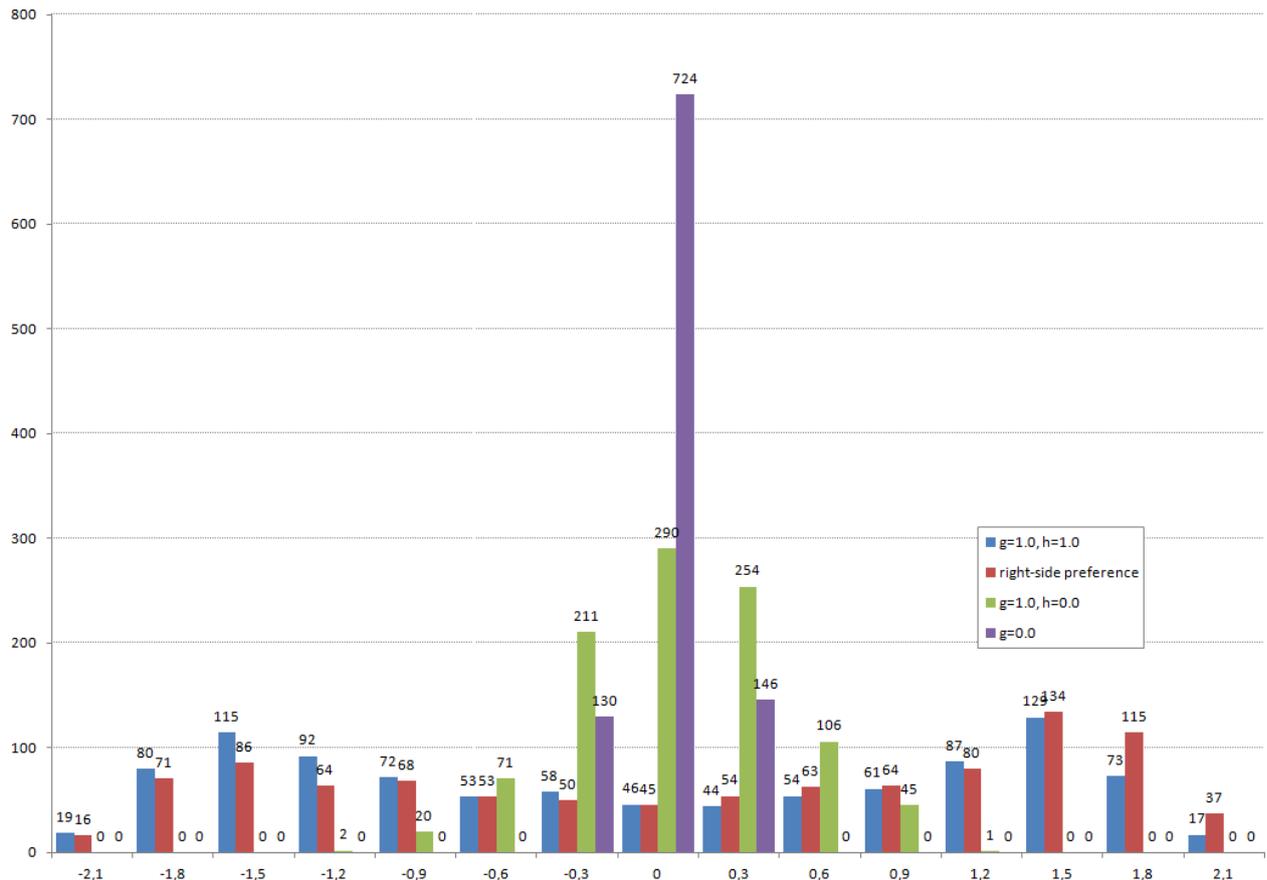

**FIGURE 8: Frequency distribution of measure *A* for four different parameter settings. Blue is the main case discussed in this contribution, i.e. the histogram is a coarser-grained version of FIGURE 5. Red is with right side preference, green is with *g=1.0*, but *h=0.0* and purple is with *g=0.0*.**